\documentstyle[12pt]{article}
\begin{document}

\baselineskip=20pt

\title{Invariance of reduced density matrices under Local Unitary operations}

\author{A. M. Martins\\
Instituto Superior T\'{e}cnico, 1049-001
Lisboa, Portugal }

\date{}

\maketitle

\begin{abstract}

ola

We derive necessary and sufficient conditions for local unitary (LU) operators to leave invariant the set of 1-qubit reduced density matrices of a multi-qubit state. LU operators with this property are tensor products of {\it cyclic local} operators, and form a subgroup, the centralizer subgroup of the set of reduced states, of the Lie group $SU(2)^{\otimes n}$. The dimension of this subgroup depends on the type of reduced density matrices. It is maximum when all reduced states are maximally mixed and it is minimum when none of them is maximally mixed. For any given multi-qubit state, pure or mixed, we compute the LU operators that fix the corresponding reduced density matrices and determine the equivalence class of the given state. \\\\

PACS number(s) 03.67.Mn, 03.65.Aa, 03.65.Ud.

\end{abstract}

\newpage

\newpage  

\section{Introduction}

Measuring and classifying quantum entanglement  has been the object of extensive research work. The motivations are related to applications in quantum information and computation tasks \cite{Bennett1992, Bennett1993} as well as to the foundations of quantum physics \cite{Ariano2010, Pusey2012}. An exhaustive bibliography about these different aspects can be found in a recent review article by Horodecki and al. \cite{Horodecki2009}

A very fruitful approach to understand entanglement, was launched by the seminal work of Linden and al. \cite{Popescu1997, Popescu1998} who first used group-theoretic methods to classify entanglement in multi-qubit systems through their classes of local unitary (LU) equivalent states. Two quantum states that can be transformed into each other by LU operations, have the same amount of entanglement and are characterized by local polynomial invariants  \cite{Grassl1998, Sudbery2001}. 

Among all possible local unitary operations that can be applied to a subsystem of a quantum system, there are the {\it cyclic operations} \cite{Fu2005}, that fix the corresponding reduced state.  These operations originate nonlocal effects in the global quantum state of the system that may distinguish product states from classically correlated states. Based on these operations new entanglement measures have been proposed  \cite{Gharibian2008, Illuminati2011}. LU operations that fix reduced states, leave also invariant local measurements (LM). We say that two quantum states are LM-equivalent when they have the same set of 1-party reduced density matrices.

In this work we answer the following question: Given an $n$-qubit input state $\rho$, pure or mixed, what is the set of states to which it can be converted by LU operations that leave invariant the corresponding reduced states? This is, what is the set of states that are LU and LM equivalent to $\rho$?

We say that a state, $\rho_U$ is LU-equivalent to $\rho$ if $\rho_U =U \rho U^{\dag}$, with $U= \otimes_j^{n} U_j ( \in G )$, where $U_j \in SU(2)$, is the unitary operator acting in qubit  $j$ and $G= SU(2)^{\otimes n}$, is the local unitary group. Each equivalence class of LU-equivalent states is an orbit of this group. We say that $\rho_U$ is LM-equivalent to $\rho$ when their set $S$, of reduced states, is the same, i.e, ${\bf T}_{(j)} (\rho_U) = {\bf T}_{(j)} (\rho) =\rho_j \,\ ( i=1,..,n)$, where ${\bf T}_{(j)}$ is the partial trace over all qubits except qubit $j$. The set of local operators $U $ that fix each of the $n$ reduced states $\rho_j$, is the centralizer subgroup of the set $S$.

We derive necessary and sufficient conditions for an LU operator to belong to the centralizer subgroup of the set $S$ and identify all possible types of centralizers subgroups. We show that their dimension is directly related with the number of maximally mixed 1-qubit states. We also prove that the operator $U_i$ that fixes any non-maximally mixed reduced state $\rho_i$, is a 1-parameter unitary operator completely determined by the Bloch vector of $\rho_i$.

The partial trace operator play a central role in the derivation of the above mentioned results and deserve a place of their own right in this work. We explore the isomorphism existent between the orthogonal complement of the kernel of ${\bf T}_{(j)}$ and the Hilbert space of qubit $j$, to identify the vectors representing the LM-equivalence classes. 

The paper is organized as follows. In Section 2, we use the partial trace operator to decompose the Hilbert space of the whole system in pairs of complementary subspaces. In Section 3, we define an  isomorphism between the reduced density matrices and vectors of the Hilbert space of the $n$-qubits and derive the necessary and sufficient conditions obeyed by a local unitary operator that fix the corresponding reduced state. In Section 4, we compute all possible centralizer subgroups of a set of reduced states and give the explicit form of the LU/LM operators. Finally we conclude in Section 5.

\section{The partial trace }

A suitable choice of the basis set  to develop the density matrices may simplify considerably solving specific physical problems, or may help to identify new properties of the system. In this work, where systems are formed by $n$ similar $2$-level constituents, and where the partial trace operators play a determinant role, the natural choice of basis set is the generalized Bloch vector basis. 

Let ${\cal V}_j$ denote the $4$-dimensional Hilbert space of $2 \times 2$ Hermitian matrices. A convenient basis for ${\cal V}_j$ is ${\cal  B}_j= \{\sigma_{\alpha_j} ; {\alpha_j}=0,1,2,3 \}$, where  $ \sigma_{\alpha_j}( \alpha_j=1,2,3)$ represents the usual Pauli matrices, and $\sigma_0 = {\bf 1}$, is the $2 \times 2$ identity matrix. Using in ${\cal V}_j$ the Hilbert-Schmidt inner product  $(\sigma_{\alpha_i }, \sigma_{\alpha_k}) = Tr\{ \sigma_{\alpha_i }\sigma_{\alpha_k } \} = 2 \delta_{ij} $, then ${\cal  B}_j $ is an orthogonal basis set. We are going to consider the set $ {\cal B}_{{\cal V}^{\otimes n}} = \{ \sigma_{\vec \alpha} \}$, where
\begin{equation}\label{vector2}
 \sigma_{\vec \alpha} = \otimes_{j=1}^{n}  \sigma_{\alpha_j}
\end{equation} 
The vector index ${\vec \alpha} =(\alpha_1, \alpha_2,... ,\alpha_n)$ is a $n$-tuple containing the $n$ indices $\alpha_j$. There exist $4^{n}$ such matrices all being traceless, except for $\sigma_{\vec 0 }= {\otimes}_{j=1}^n {\bf 1}_j$, which corresponds to the $2^n \times  2^n $ identity matrix with trace $Tr \{ \sigma_{\vec 0 } \} = 2^{n}$.

$ {\cal B}_{{\cal V}^{\otimes n}} $ is an orthogonal basis set of the complex $4^{n}$-dimensional Hilbert-Schmidt vector space ${\cal V}^{\otimes n} =\otimes_{j=1}^n {\cal V}_j $. Every complex square matrix, $(2^{n} \times 2^{n})$, can be seen as a vector $\bf v$, uniquely written in the form
\begin{equation}\label{vector}
{\bf v} =  \sum_{\vec \alpha} v_{\vec \alpha} \,\ \sigma_{\vec \alpha}
\end{equation}
where the components $v_{\vec \alpha} $ are given by 
\begin{equation}\label{components}
v_{\vec \alpha} =  \frac{1}{2^n} Tr\{ \sigma_{\vec \alpha} \,\ {\vec v}  \}
\end{equation}

Any $n$-qubit quantum state $\rho=\sum_{\vec \alpha} r_{\vec \alpha}  \sigma_{\vec \alpha} \in {{\cal V}^{\otimes n}} $, must be hermitian, $\rho = \rho^{\dag}$, definite positive $\rho \geq 0$, and normalized $Tr \{ \rho \} =1$. These requirements on $\rho$ impose certain constrains to the components $ r_{ \vec \alpha} $: (a) $\forall_{\vec \alpha}, r_{\vec \alpha} \in \Re$,  (b) $r_{\vec 0} = \frac{1}{2^n} $, (c) $r_{\vec \alpha} = \frac{1}{2^n} Tr\{ \sigma_{\vec \alpha} \,\ \rho \}$ and (d) $\sum_{ \vec \alpha} r_{ \vec \alpha}^{2} \leq 1 $, the equality is attained for pure states.

The translated vector, ${\bar \rho }= \rho -  {\bf 1}^{ \otimes n} /2^{n}$,( ${\bf 1}^{ \otimes n} = \otimes_{j=1}^{n}  {\bf 1}_j$),  characterizes completely the quantum state $\rho$ and is the well known {\it generalized Bloch vector representation} of dimension $(4^{n} - 1)$.

Let $D_n$ be the set of the $n$-qubit density matrices $\rho$ and let ${\bf T}_{(i)} : {\cal V}^{\otimes n} \rightarrow {\cal V}_i $ be the linear transformation defined by
\begin{equation}\label{partial}
{\bf T}_{(i)} (\rho)  =Tr_{n/  \{i  \}} \{ \rho \} = \rho_i \,\,\,\,\,\,\,\,\,\,\,\,\  ( i=1,...,n)
\end{equation}
where $Tr_{n/  \{i  \}} \{ . \}$ is the partial trace operator over $(n-1)$ qubits, except qubit $i$. 

This is a surjective map of ${\cal V}^{\otimes n}$, (the vector space of the $n$ qubits), onto ${\cal V}_i \equiv Im({\bf T}_{(i)}  ) $, (the vector space of qubit $i$), where $Im({\bf T}_{(i)} )$ is the image space of ${\bf T}_{(i)} $. Let $ {\cal K}_i $ be the kernel of ${\bf T}_{(i)} $ and let ${\cal Q}_i$ be the orthogonal complement of $ {\cal K}_i $, i.e.,
\begin{equation}\label{partial}
{\cal V}^{\otimes n} = {\cal Q}_i \oplus {\cal K}_i   \,\,\ ,\,\,\,\,\,\,\,\,\,\,\,\,\,\,\,\   \,\,\ (i=1,...,n)
\end{equation}
The subspace $ {\cal Q}_i $ is a $4$-dimensional space isomorphic to ${\cal V}_i  $. Applying the map ${\bf T}_{(i)}  $ to the vectors of the basis set
\begin{equation}\label{basis2}
 {\cal  B}_{{\cal Q}_i} = \{ {\bf b}_{\alpha_i}= \otimes_{k=1}^{i-1}  {\bf 1}_k  \otimes \sigma_{\alpha_i}  \otimes_{k'=i+1}^n {\bf 1}_{k'} ; \,\   \alpha_i = 0,1,2,3\}  
\end{equation}
 where $ {\cal  B}_{{\cal Q}_i} \subset {\cal B}_{{\cal V}^{\otimes n}} $, we obtain 
\begin{equation}\label{correspondance}
{\bf T}_{(i)} ({\bf b}_{\alpha_i} )=2^{n-1} \sigma_{\alpha_i}  
\end{equation}
Let ${\cal  B}_{{\cal K}_i} =  {\cal B}_{{\cal V}^{\otimes n}} \setminus  {\cal  B}_{{\cal Q}_i}  $. The image of any vector of ${\cal  B}_{{\cal K}_i} $ under ${\bf T}_{(i)}$ is the zero vector $0_i \in {\cal V}_i $. We conclude that ${\cal  B}_{{\cal Q}_i}$ and ${\cal  B}_{{\cal K}_i}$ are orthogonal basis sets for the subspaces ${\cal Q}_i$ and ${\cal K}_i$, such that, ${\cal B}_{{\cal V}^{\otimes n}} = {\cal  B}_{{\cal Q}_i}  \cup {\cal  B}_{{\cal K}_i}  $. 

Any density matrix $\rho \in D_n$ can be written in a unique way as
\begin{equation}\label{direct}
 \rho =\rho_{{\cal Q}_i}  + \rho_{{\cal K}_i} 
 \end{equation}
where $\rho_{{\cal Q}_i}$ and $ \rho_{{\cal K}_i} $ are the projections of $\rho$ on the subspaces ${\cal Q}_i $ and $ {\cal K}_i $.

The projection operator $\pi_{{\cal Q}_i} : {\cal V}^{\otimes n} \rightarrow {\cal Q}_i$ is defined by
\begin{equation}\label{rhoq1}
\pi_{{\cal Q}_i} (\rho) = \rho_{{\cal Q}_i} = \sum_{{\alpha_i}=0}^3 \frac{(\rho , {\bf b}_{\alpha_i})}{\parallel {\bf b}_{\alpha_i} \parallel^2 } {\bf b}_{\alpha_i}= \sum_{{\alpha_i}=0}^3 r_{0_1 ... 0_{i-1} \alpha_i 0_{i+1}...0_n}{\bf b}_{\alpha_i}
 \end{equation}
The action of ${\bf T}_{(i)}$ on both sides of eq.(\ref{direct}), gives
\begin{equation}\label{rhoq2}
{\bf T}_{(i)}(\rho ) ={\bf T}_{(i)}(\rho_{{\cal Q}_i} )=  \rho_i
\end{equation}
Any 1-qubit density matrix $\rho_i \in {\cal V}_i$ can be written in the basis ${\cal B}_i $ in the form
\begin{equation}\label{rhoi}
\rho_i  =  \frac{1}{2} ({\bf 1}_i+ {\vec r}_i . {\vec \sigma}(i))  \,\,\,\,\,\,\,\,\,\,\,\,\  ( i=1,...,n)
\end{equation}
where, $ {\vec r}_i . {\vec \sigma}(i) =  2^n \sum_{a_i =1}^3  r_{0_1 ... 0_{i-1} a_i 0_{i+1}...0_n}  \sigma_{a_i}$, and ${\vec r}_i $ is the Bloch vector of qubit $i$, such that
$\| {\vec r}_i  \|  \leq 1  $. 

Substituting (\ref{rhoi}) in the l.h.s. of eq.(\ref{rhoq1}), we obtain the following explicit one-to-one correspondence between the vectors $\rho_i \in {\cal V}_i $ and the vectors $\rho_{{\cal Q}_i} \in {\cal Q}_i$,
\begin{equation}\label{rhoq5}
\rho_{{\cal Q}_i} = \frac{1}{2^{n-1}} \otimes_{k=1}^{i-1}  {\bf 1}_k  \otimes \rho_i  \otimes_{k'=i+1}^n {\bf 1}_{k'} 
 \end{equation}
 
The translated vector
\begin{equation}\label{rhoqj2}
{\bar  \rho}_{{\cal Q}_i} = \rho_{{\cal Q}_i} - \frac{1}{ 2^n}{\bf 1}^{ \otimes n}=  \otimes_{k=1}^{i-1}  {\bf 1}_k  \otimes [  {\vec r}_i . {\vec \sigma}(i)]  \otimes_{k'=i+1}^n {\bf 1}_{k'} 
\end{equation}
contains the same quantum information as the reduced density matrix $\rho_i$, this is, the quantum state of qubit $i$, is fully represented by the vector ${\bar  \rho}_{{\cal Q}_i} \in {\cal V}^{\otimes n}$. Moreover, 
\begin{equation}\label{orto}
( {\bar  \rho}_{{\cal Q}_i} ,{\bar  \rho}_{{\cal Q}_j}  ) = \| {\bar  \rho}_{{\cal Q}_i}  \|^2 \delta_{ij}
\end{equation}
 i.e., vectors ${\bar  \rho}_{{\cal Q}_i} $
associated to different qubits are orthogonal to each other and to any other vectors of the basis set $ {\cal B}_{{\cal V}^{\otimes n}} $. The norm, $\| {\bar  \rho}_{{\cal Q}_i}  \| = [ \lambda_{-}^2 (i) +\lambda_{+}^2 (i) ]^{1/2} $, where, $\lambda_{\mp}$, are the eigenvalues of ${\bar  \rho}_{{\cal Q}_i}$. It is now obvious that any quantum state $\rho  \in D_n$ can be written in the form,
\begin{equation}\label{rho4}
 \rho = \frac{1}{2^n} {\bf 1}^{ \otimes n}   + \sum_{i=1}^{n} {\bar  \rho}_{{\cal Q}_i}  + \Delta 
\end{equation}
where $\Delta$ refers to the terms of $\rho$ containing all possible $k$-partite correlations ($ 2 \leq k \leq n $) existing between the $n$-qubits. 

\section{ Reduced states and equivalence classes}

The possible outcomes of the measurement of any local observable $ {\hat A}_j  \in {\cal V}_j $, performed on qubit $j$, are given by the eigenvalues $a_k(j)$ of the operator ${\hat A}_j $. The expectation value of this measurement, when the system is in the state $\rho$, is given by 
\begin{equation}\label{expectation}
\langle {\hat A}_j  \rangle  = Tr \{{\hat A}_j  \rho \} = Tr_j \{ {\hat A}_j  \rho_j \}
\end{equation} 
where $Tr_j \{ \} $ is the trace in qubit $j$ and $Tr \{  \}$ is the trace in all qubits. This equality shows that measurements performed on qubit $j$ give the same result as if it would be in the reduced state $\rho_j =  {\bf T}_{ (j) } (\rho)  $. A imediate consequence of eq.(\ref{expectation}) is that different global quantum states $\rho$ with equal reduced states $  \rho_j $ have equal $1$-qubit expectation values $\langle {\hat A}_j  \rangle $.  When two states $\rho $ and $\rho^{'} $ have the same image $\rho_j$, under the map ${\bf T}_{ (j) }$, their difference belong to the kernel ${\cal K}_j$, i.e., they are congruent modulo ${\cal K}_j$. The set of all states with reduced state $\rho_j$ forms a LM$_j$-equivalence class $ C_j$, this is,
\begin{equation}\label{class}
C_j = \{ \rho \in D_n : {\bf T}_{ (j) } (\rho) =\rho_j \}
\end{equation}
The set of all LM$_j$-equivalence classes is the quotient space ${\cal V}^{\otimes n}/ {\cal K}_j$. 

We have shown that for any quantum state $\rho$, there is a one-to-one correspondence between its reduced state $\rho_j$ and its projection $\rho_{{\cal Q}_j }$. This enables us to define a linear map $\psi_j: {\cal V}^{\otimes n}/ {\cal K}_j \rightarrow {\cal Q}_j $, such that
\begin{equation}\label{class}
\psi (C_j )= \rho_{{\cal Q}_j }
\end{equation}
assigns to each class $C_j \in {\cal V}^{\otimes n}/ {\cal K}_j$ the vector $\rho_{{\cal Q}_j } \in {\cal Q}_j$, we say that the vector $\rho_{{\cal Q}_j } \in {\cal Q}_j$, is the representative state of the class $C_j $ and we may write \cite{Martins2008}
\begin{equation}\label{class1}
C_j = \{ \rho \in D_n : \rho =\rho_{{\cal Q}_j } + {\cal K}_j  \}
\end{equation}
The set of all $n$-qubit density matrices, such that their reduced density matrices belong to $S= \{ \rho_i = {\bf T}_{ (i) } (\rho), \,\ ( i=1,...,n )  \}$, is given by the intersection of the equivalence classes $C_i$, i.e.,
\begin{equation}\label{class2}
{\bar C} = \bigcap_{i=1}^{n} C_i = \{ \rho \in D_n : \rho =\rho_{{\cal Q}_i } + {\cal K}_i \,\ ; i=1,...,n \}
\end{equation}
saying it in another way, quantum states in the set ${\bar C}$ have their $\rho_{{\cal Q}_i}$ projections in the set  
\begin{equation}\label{set3}
{\bar S} = \{ \rho_{{\cal Q}_i}= \pi_{{\cal Q}_i}(\rho) \,\ ; i=1,...,n   \}
\end{equation} 
The set ${\bar S}$ is isomorphic to the set $S$ therefore, they have the same content of quantum information. This isomorphism is particularly useful when we are studying local properties of the qubits because, instead of working with the $n$ Hilbert spaces ${\cal V}_i$, we can use the original Hilbert space ${\cal V}^{\otimes n}$ of the $n$-qubits.

The unitary transformation $U  \in G$ acts on a $n$-qubit state $\rho$ via the adjoint action,
\begin{equation}\label{LU}
\rho_U = ad \,\ U [ \rho] = U \rho U^{\dag} = \left( \otimes_{j=1}^{n} U_j \right) \rho  \left( \otimes_{j=1}^{n} U_j^{\dag} \right) 
\end{equation}
where $G= SU(2)^{\otimes  n}$ is a $3n$-dimensional Lie group and ${\cal L} = su(2) \oplus su(2) \oplus  ... \oplus su(2) $ is the corresponding Lie algebra. The set ${\cal B}_{\cal L} =\{  {\bf b}_{a_i} \in {\cal B}_{{\cal Q}_i}; \,\ a_i =1,2,3 $ and $ i=1,...,n \}$ is a basis set for ${\cal L}$ whose elements are the generators of $G$. 

In this work we are looking for all quantum states $\rho_U $, LU equivalent to $\rho$, such that measurements of any local observable ${\hat A}_i$ are not able to distinguish between $\rho$ and $\rho_U$. Having in mind eq.(\ref{expectation}), we are looking for states $\rho_U $ with the same set $S$ of 1-qubit reduced density matrices. This is,
\begin{equation}\label{trace2}
{\bf T}_{(i)} (\rho_U) = {\bf T}_{(i)} (\rho) =\rho_i  ; \,\,\,\,\ i=1,...,n
\end{equation}
or, given the one-to-one correspondence between $\rho_i$ and $\rho_{{\cal Q}_i}$, the LU equivalent states are such that
\begin{equation}\label{trace3}
\pi_{{\cal Q}_i} (\rho_U) =  \pi_{{\cal Q}_i} (\rho) = \rho_{{\cal Q}_i} ; \,\,\,\,\ i=1,...,n
\end{equation}
i.e., the states $ \rho_U $ belong to the set ${\bar S}$. Not all adjoint actions of local unitary operators $U$ on $\rho$ obey this condition, however all local unitary operators leave the subspaces ${\cal Q}_j$ and ${\cal K}_j$ invariant, as we prove in the next Theorem. 

{\bf Theorem 1:} {\it The subspaces ${\cal Q}_j$ and ${\cal K}_j$ are invariant under LU transformations.}

{\bf Proof:}  Any vector ${\bf v}_{{\cal Q}_j} \in {\cal Q}_j$ has the form ${\bf v}_{{\cal Q}_j} = \otimes_{k=1}^{i-1}  {\bf 1}_k  \otimes [ \sum_{\alpha_i=0}^{3}v _{\alpha_i }  \sigma_{\alpha_i} ]  \otimes_{k'=i+1}^n {\bf 1}_{k'} $. The adjoint action of $U $ on $\rho$ is
\begin{equation}\label{invariant}
ad \,\ U [ {\bf v}_{{\cal Q}_j} ] = U {\bf v}_{{\cal Q}_j}  U^{\dag} =\otimes_{k=1}^{i-1}  {\bf 1}_k  \otimes [ \sum_{\alpha_i=0}^{3}v _{\alpha_i } U_i  \sigma_{\alpha_i} U_i^{\dag} ] \otimes_{k'=i+1}^n {\bf 1}_{k'} 
\end{equation}
As 
\begin{equation}\label{rotation}
\sum_{\alpha_i=0}^{3}v _{\alpha_i } U_i  \sigma_{\alpha_i} U_i^{\dag}  =  \sum_{\alpha_i=0}^{3}v _{\alpha_i }^{'}  \sigma_{\alpha_i}
\end{equation}
 then $ U {\bf v}_{{\cal Q}_j}  U^{\dag} \in  {\cal Q}_j$. 
 
 When ${\cal Q}_j$ (or ${\cal K}_j$ ) is invariant under a unitary transformation so is the complementary subspace ${\cal K}_j$ (or ${\cal Q}_j$) \cite{Halmos1987}. $\Box$
 
 {\bf Corollary 1:} {\it The subspace ${\bar {\cal K}} = \cap_{j=1}^{n} {\cal K}_j$ is invariant under the local adjoint action.}

 {\bf Corollary 2:} {\it The projection operator $\pi_{{\cal Q}_i}$ commutes with any LU transformation, i.e.,}
 \begin{equation}\label{comutador2}
\pi_{{\cal Q}_i} (U  \rho U^{\dag})  = U [ \pi_{{\cal Q}_i} ( \rho ) ] U^{\dag}  = U \rho_{{\cal Q}_i}  U^{\dag}
\end{equation}
{\bf Proof: }
If a subspace is invariant under a linear transformation $U$ then $U$ commutes with every projection operator on that subspace \cite{Halmos1987}. $\Box$

This corollary shows that $\pi_{{\cal Q}_i} (  \rho_U) =U \rho_{{\cal Q}_i}  U^{\dag}$. Imposing now the constrain of eq.(\ref{trace3}), i.e., that $ \rho_U $ has the same set of 1-qubit reduced density matrices as $\rho$, we conclude that the LU transformations we are looking for, are such that
\begin{equation}\label{inv}
U \rho_{{\cal Q}_i}  U^{\dag} = \rho_{{\cal Q}_i} \,\,\ ; \,\,\,\ i=1,...,n
\end{equation} 
$\rho_{{\cal Q}_i}$ is invariant under LU transformations. Local unitary operators $U \in G$ obeying condition (\ref{inv}), for all elements of the set ${\bar S} $, belong to the centralizer subgroup $C_G ({\bar S})$ of the set ${\bar S}$, i.e.
\begin{equation}\label{centralizergroup}
C_G ({\bar S}) = \{U \in G : U  \rho_{{\cal Q}_i} U^{\dag} = \rho_{{\cal Q}_i}, \forall_{\rho_{{\cal Q}_i} \in {\bar S}}  \}
\end{equation}

Next theorem sets the conditions obeyed by the local unitary transformations $U_i $ in order that equality (\ref{inv}) holds. 

{\bf Theorem 2 :} {\it A state $\rho_U $, LU equivalent to $\rho$, has the same set $S$ of reduced density matrices as $\rho$, iff each local unitary operator $U_i \in SU(2)$ commutes with $\rho_i$, i.e.,}
\begin{equation}\label{comutador}
[U_i , \rho_i]=0 \,\,\ , \,\,\    i=1,...,n
\end{equation}

{\bf Proof: } By Corollary 1, 
\begin{equation}\label{trans}
\pi_{{\cal Q}_i} (U  \rho U^{\dag}) =U \rho_{{\cal Q}_i}  U^{\dag} =  \otimes_{k=1}^{i-1}  {\bf 1}_k  \otimes  U_i  \rho_i U_i^{\dag} \otimes_{k'=i+1}^n {\bf 1}_{k'}
\end{equation}
The condition (\ref{inv}) is verified when, $U_i  \rho_i U_i^{\dag}  = \rho_i $ for each qubit $i$. This is equivalent to equality (\ref{comutador}).       $  \Box$

Theorem 2 refers to these multi-qubit LU operations and proves that the cyclic property is a necessary and sufficient condition for invariance of any number of reduced states.  Moreover, local unitary operators acting in different qubits $i$ and $j$, commute with each other, i.e., $ [ U_i , U_j ] = 0$.

In conclusion, the general form of any quantum state $\rho_U$, LU equivalent to $\rho$, and with the same set $S$ of 1-qubit reduced density matrices is given by
\begin{equation}\label{flu}
 \rho_U = \frac{1}{2^n} {\bf 1}^{ \otimes n}   + \sum_{i=1}^{n} {\bar  \rho}_{{\cal Q}_i}  + U \Delta U^{\dag}
\end{equation}
where the operators $U$ belong to the centralizer subgroup $C_G ({\bar S})$. The problem of finding $\rho_U$ in the last equation is solved when the centralizer subgroup of a state $\rho$ is known.

We call $\rho $-family and denote by ${\cal F}_{\rho}$, the set of states $\rho_U$ given by eq.(\ref{flu}). The elements of this family have the same type of entanglement but are not distinguishable by local measurements.

 Not all states in the LU-orbit of $\rho$ belong to ${\cal F}_{\rho}$. Next proposition is a criterium to decide wether a state $\rho^{'}$, is not in the family ${\cal F}_{\rho}$.

{\bf Proposition 1:} A state $\rho^{'}$, LM-equivalent to the state $\rho$, does not belong to the family  ${\cal F}_{\rho}$, if 
\begin{equation}\label{cond}
Tr\{  \rho^{' 2} \} \neq  Tr\{ \rho^{2} \}   \,\,\,\,\ \mbox{or if}  \,\,\,\,\  Tr\{ \Delta^{' 2} \} \neq  Tr\{ \Delta^{2} \}  
\end{equation}

\section{Centralizers subgroups and LU/LM-equivalence}

In this section we show that the translated vectors $ {\bar  \rho}_{{\cal Q}_i}$, present in eq.(\ref{rho4}), determine the centralizer subgroup $C_G ({\bar S})$ and the set of states LU/LM equivalent to each quantum state $\rho$.

Any generic local unitary operator $U_j \in SU(2)$ is a three real continuous parameter operator and can be written in the form
\begin{equation}\label{uni}
U_j (  \phi_j  , \theta_j , \omega_j )=  e^{i {\vec s}_j .{\vec \sigma} (j) } = \cos ( \omega_j ) {\bf 1}_j +i \sin (\omega_j ) {\hat n}_{ {\vec s}_j} . {\vec \sigma} (j)
\end{equation}
where ${\hat n}_{ {\vec s}_j} = {\vec s}_j /  \parallel  {\vec s}_j    \parallel  \equiv ( \cos \phi_j  \sin \theta_j ,  \sin \phi_j  \sin \theta_j, \cos \theta_j )$ is a unit vector in the 3-dimensional Euclidian space (Bloch space of qubit $j$), parametrized by the azimuthal angle, $0 \leq \phi_j \leq 2 \pi $, and the polar angle, $0 \leq \theta_j \leq \pi $. The third parameter is $\omega_j =  \parallel  {\vec s}_j  \parallel$ ($ 0 \leq \omega_j  \leq \pi /2$) is the length of the vector ${\vec s}_j$.

Any 1-qubit density matrix can be written in the form (\ref{rhoi}). When ${\vec r}_j  =0$, then the 1-qubit density matrix is maximally mixed, i.e., $\rho_j^{*} =\frac{1}{2}{\bf 1}_j $, and any local unitary operation $U_j = e^{i {\vec s}_j .{\vec \sigma} (j) }$
commutes with $\rho_j$. 
 
When ${\vec r}_j  \neq 0$, then condition (\ref{comutador}) is verified when ${\vec s}_j  = \xi_j {\vec r}_j $ (see Appendix), with $\xi_j \in \Re$. The corresponding {\it local cyclic} operator is 
\begin{equation}\label{uni2}
U_j (\xi_j )= e^{i \xi_j {\vec r}_j .{\vec \sigma} (j) } = \cos ( \omega_j ) {\bf 1}_j + i \sin (\omega_j ) {\hat n}_{ {\vec r}_j} . {\vec \sigma} (j)
\end{equation}
a single parameter unitary operator, where $ \omega_j  = \xi_j   \parallel  {\vec r}_j    \parallel $ is the continuous parameter. The direction ${\hat n}_{ {\vec r}_j}$ is fixed by the cyclic condition (\ref{comutador}). Varying continuously the parameter $\xi_j $, in eq.(\ref{uni2}), between, $0$, and, $\pi /2  \parallel  {\vec r}_j    \parallel $, then $U_j $ varies between $ {\bf 1}_j $  and $ ( i {\hat n}_{ {\vec r}_j} . {\vec \sigma} (j) )$.

Invoking the local isomorphism between SU(2) and SO(3) we see that the unitary operator $U_j$, of eq.(\ref{uni2}), represents a rotation of an angle $\omega_j$ around the vector ${\vec r}_j $ of the Bloch sphere of qubit $j$, which leaves this vector and the corresponding $\rho_j$ invariants. In the generalized $(4^{n}-1)$ Bloch vector space, the vectors ${\bar  \rho}_{{\cal Q}_j}$, of different qubits, are orthogonal to each other. Local unitary operations of $SU(2)^{\otimes n}$ of the type
\begin{equation}\label{uni3}
U = \otimes_{j=1}^{m} e^{i \xi_j {\vec r}_j .{\vec \sigma} (j) } \otimes_{l=m+1}^{n} {\bf 1}_l \,\,\,\, (m=1,...,n)
\end{equation}
correspond to $m$ independent rotations of the the group $SO(3)$ around each vector ${\bar  \rho}_{{\cal Q}_j}$.

These results reveal the intimate relation between the set $S$ of the 1-qubit reduced density matrices and the centralizer subgroup $C_G ({\bar S})$ of the set ${\bar S}$. Different cases are possible.

{\bf Case 1:} When $ {\bar  \rho}_{{\cal Q}_i} =0 $, for all $i=1,...,n$, then eq.(\ref{rho4}) reduces to
\begin{equation}\label{rho5}
 \rho = \frac{1}{2^n} {\bf 1}^{ \otimes n}   + \Delta 
\end{equation}
and $\rho$ have maximally mixed 1-qubit reduced states, i.e.,
\begin{equation}\label{set1}
S= \{ \rho_i^{*}= \frac{1}{2} {\bf 1}_i \,\ ; \,\  i=1,...,n\}
\end{equation}
The centralizer subgroup $C_G ({\bar S})$ is the entire $G$ whose dimension is $3n$. The states $\rho_U $, are given by
\begin{equation}\label{rho6}
 \rho_U = \frac{1}{2^n} {\bf 1}^{ \otimes n}   + U \Delta U^{\dag} 
\end{equation}
where $ U =\otimes_{j=1}^{n} e^{i{\vec s}_j .{\vec \sigma} (j) } $. As each $U_j$ only acts on qubit $j$ then $U \Delta U^{\dag}$ can be easily computed by replacing each Pauli matrix $\sigma_{\alpha_j}$ in $\Delta$ by $U_j \sigma_{\alpha_j} U_j^{\dag}$.

Any n-qubit Werner state $\rho^{\cal W}$, has maximally mixed 1-qubit reduced states. All states in the LU-orbit of a $\rho^{\cal W}$ are LM-equivalent. The maximally mixed state $\rho^{*}= \frac{1}{2^n} {\bf 1}^{ \otimes n} $ is a special type of Werner state.  When all local unitary operators are equal, i.e., $U_j =e^{i{\vec s} .{\vec \sigma} (j) }$ (independent o $j$) the state $\rho^{\cal W}$ is transformed into itself.  

The $n$-GHZ entanglement class has maximally mixed 1-qubit density matrices. All states in this class are LM-equivalent.

{\bf Case 2:} When, in eq.(\ref{rho4}), there are $m < n$ operators $ {\bar  \rho}_{{\cal Q}_k} \neq 0, (k=1,...,m) $ and the remaining $(n-m)$ operators are $ {\bar  \rho}_{{\cal Q}_l} =0, (l=m+1,...,n) $, then
\begin{equation}\label{rho7}
 \rho = \frac{1}{2^n} {\bf 1}^{ \otimes n}   + \sum_{i=1}^{m} {\bar  \rho}_{{\cal Q}_i}  + \Delta 
\end{equation}
The corresponding 1-qubit reduced density matrices belong to the set 
\begin{equation}\label{set2}
S=\{  \rho_k =\frac{1}{2}{\bf 1}_k + {\vec r}_k . {\vec \sigma} (k) ; \,\ ( k=1,...,m)  \wedge   \rho_l^{*} =\frac{1}{2}{\bf 1}_l ;  ( l=m+1,...,n) \}
\end{equation}
The centralizer subgroup $C_G ({\bar S})$ is 
\begin{equation}\label{centralizergroup2}
C_G ({\bar S}) = \{U \in G : U = \otimes_{k=1}^{m}e^{i\xi_l {\vec r}_l .{\vec \sigma} (l) }   \otimes_{l=m+1}^{n}  e^{i {\vec s}_k .{\vec \sigma} (k) } \}
\end{equation}
Different values of $m$ give rise to different centralizer subgroups with dimension $dim[ C_G ({\bar S})] =3n -2m$, the same as the number of independent continuous parameters. When $m=n$, $dim[ C_G ({\bar S})] =n$, this number is minimum. When $m=0$, then $dim[ C_G ({\bar S})] =3n $ is maximum and the centralizer subgroup is the entire $G$ (Case 1). 

The states $\rho_U $, of the family  ${\cal F}_{\rho}$ are obtained by replacing $\Delta$ in eq.(\ref{rho7}) by $U \Delta U^{\dag}$, where $U $ belong to the stabilizer subgroup of eq.(\ref{centralizergroup2}).

Biseparable states 
\begin{equation}\label{by}
 \rho = \rho^{(m)} \otimes \rho^{(n-m)}
\end{equation} 
where $\rho^{(m)}$ is any state of $m$-qubits and the remaining $(n-m)$-qubits are in a state $\rho^{(n-m)}$, (for instance, $(n-m)$-GHZ or $(n-m)$-Werner states), have centralizers subgroups of the form (\ref{centralizergroup2}). A product state 
\begin{equation}\label{product}
 \rho = \otimes_{k=1}^{m} \rho_k \otimes_{l=m+1}^{n} \rho_l^{*}
\end{equation}
is a special case of biseparable state. The $\rho$-family of a product state is the product state itself.

 \section{Concluding remarks}
 
 We have investigated a special type of local unitary operations that fix the set of reduced states of a pure or mixed multi-qubit state. We have shown that  the possible forms of the {\it cyclic local} transformations is determined by the 1-qubit reduced density matrices. The dimension of the centralizer subgroups of the set reduced states is minimum when no 1-qubit reduced matrix is maximally mixed and it is maximum when all 1-qubit reduced states are maximally mixed.

We have shown that local {\it cyclic} unitary operations of multiqubit states, with non maximally mixed 1-qubit reductions, are a subgroup of $SU(2)^{\otimes n}$ whose elements are given by the tensorial product of $n$-single parameter unitary operators, this suggests that it is possible to analyze in a continuous way the measures of entanglement and of non-classicality \cite{Gharibian2012}, by varying independently these parameters. Simultaneous application of these local {\it cyclic} operations to different qubits goes beyond the bipartite studies \cite{Fu2005,Gharibian2008} and may reveal new nonlocal effects.

 \section{Appendix}
 
  {\bf Theorem 3}: {\it The commutation relation $ [U_j , \rho_j ]=0 $, where the 1-qubit reduced state is $\rho_j =  \frac{1}{2} {\bf 1}_j+ {\vec r}_j . {\vec \sigma}(j)$, with ${\vec r}_j \neq 0$, is verified iff }
\begin{equation}\label{uni5}
U_j (\xi_j )= e^{i \xi_j {\vec r}_j .{\vec \sigma} (j) } = \cos ( \omega_j ) {\bf 1}_j + i \sin (\omega_j ) {\hat n}_{ {\vec r}_j} . {\vec \sigma} (j)
\end{equation}  
  
{\bf Proof}: Any local unitary operator has the general form (\ref{uni}). The cyclic condition is equivalent to  
\begin{equation}\label{comutador2}
[ { {\vec s}_j} . {\vec \sigma} (j) , {\vec r}_j . {\vec \sigma}(j) ] = 0
\end{equation}  
Computing the above commutator, we obtain
\begin{equation}\label{comutador3}
[ {\vec s}_j . {\vec \sigma} (j) , {\vec r}_j . {\vec \sigma}(j) ] = \sum_{k=1}^{3} \sum_{l=1}^{3} s_{jk} r_{jl} [ \sigma_k (j) , \sigma_l (j) ] = 2 i \{ \sum_{k=1}^{3} \sum_{l=1}^{3} s_{jk} r_{jl} \epsilon_{klu} \} Ê\sigma_r (j) 
\end{equation}    
where $\epsilon_{klu}$ is the Levi-Civita symbol. After some straightforward calculations we show that the commutator will be zero iff, ${\vec s}_j = \xi_j  {\vec r}_j $.

\end{document}